\documentclass{jpp}
\usepackage[final]{changes}

\usepackage[acronym]{glossaries}
\newacronym[plural=NNs,firstplural=Artificial neural networks (NNs)]{NN}{NN}{artificial neural network}
\newacronym[plural=MLPs,firstplural=multilayer perceptrons (MLP)]{MLP}{MLP}{multilayer perceptron}
\newacronym[]{MHD}{MHD}{Magnetohydrodynamic}
\newacronym[sort=W7-X]{W7X}{\protect\mbox{W7-X}}{Wendelstein~\protect\mbox{7-X}}
\newacronym[plural=PDEs,longplural={Partial Differential Equations}]{PDE}{PDE}{Partial Differential Equation}
\newacronym[plural=PINNs,firstplural=physics informed neural networks (PINNs)]{PINN}{PINN}{physics informed neural network}

\usepackage{booktabs}
\usepackage{amsmath}
\usepackage{bm}
\usepackage{siunitx}
\usepackage{xspace}
\usepackage[section]{placeins}
\usepackage{graphicx}
\usepackage{float}
\usepackage{caption}
\usepackage{hyperref}

\definechangesauthor[name={Timo Thun}, color=red]{TT}
\colorlet{Changes@Color}{red}
\makeatletter
\@namedef{Changes@AuthorColor}{red}
\colorlet{Changes@Color}{red}
\makeatother

\newcommand{\betavol}{\langle \pmb{\beta} \rangle_\mathrm{vol}}
\newcommand{\etap}{\eta_\mathrm{p}}
\newcommand{\etaptest}{\eta_\mathrm{p, test}}
\newcommand{\etaptrain}{\eta_\mathrm{p, train}}
\newcommand{\fvolnorm}{\langle \mathbf{F} \rangle_\mathrm{vol, norm}}
\newcommand{\desc}{\texttt{DESC}\xspace}
\newcommand{\descs}{\texttt{DESC}'s\xspace}

\newcommand{\yprojected}{\mathbf{y}}
\newcommand{\nfp}{\mathrm{N}_\mathrm{FP}}

\DeclareMathOperator*{\argmin}{arg\,min}

\usepackage{pgf}

\usepackage[utf8]{inputenc}
\usepackage[T1]{fontenc}

\shorttitle{Narrow Operator Models of Stellarator equilibria}
\shortauthor{T. Thun, R. Conlin, D. Panici, D. Böckenhoff}

\title{Narrow Operator Models of Stellarator Equilibria in Fourier Zernike Basis}

\author{Timo Thun\aff{1}
  \corresp{\email{timo.thun@ipp.mpg.de}},
 Rory Conlin\aff{2}, Dario Panici\aff{3}, Daniel Böckenhoff\aff{1}}

\affiliation{\aff{1}Max-Planck-Institute for Plasma Physics, 17491 Greifswald, Germany
\aff{2}Institute for Research in Electronics \& Applied Physics, University of Maryland, MD 20742, USA
\aff{3} Mechanical and Aerospace Engineering, Princeton University, NJ 08544, USA}

\begin{document}

\maketitle

\begin{abstract}
Numerical computation of the ideal \gls{MHD} equilibrium magnetic field is at the base of stellarator optimisation and provides the starting point for solving more sophisticated \glspl{PDE} like transport or turbulence models.
Conventional approaches solve for a single stationary point of the ideal \gls{MHD} equations, which is fully defined by three invariants and the numerical scheme employed by the solver.
We present the first numerical approach that can solve for a continuous distribution of equilibria with fixed boundary and rotational transform, varying only the pressure invariant.
This approach minimises the force residual by optimising parameters of \glspl{MLP} that map from a scalar pressure multiplier to the Fourier Zernike basis as implemented in the modern stellarator equilibrium solver \desc.
\end{abstract}

\section{Introduction}\label{sec:introduction}
Stellarators are inherently steady-state plasma confinement devices, which is among the key reasons behind their renaissance as promising candidates for fusion power plants.
Ideal \gls{MHD} equilibria are a central part in optimising the complex, three-dimensional plasma shapes which are a necessary condition for steady-state operation of such devices.\newline
The equilibrium magnetic field is required not only in optimisation, but also plays a role in future real-time control algorithms and simulation frameworks~\citep{Schissel2025}.

Solving the three-dimensional \gls{MHD} equations requires numerical approaches, because no analytical solution throughout the full volume of \added{strongly shaped toroidal} ideal \gls{MHD} equilibria with nested magnetic topology exists yet~\citep{Bruno1996}.
Recent work advanced analytical models for Fourier components of the equilibrium magnetic field in a subset of reactor-relevant magnetic fields and analytical expansions close to the magnetic axis are used extensively in research~\citep{Nikulsin2024, Sengupta2024}.
These analytical solutions and the following numerical solvers assume nested magnetic topology, or integrability throughout the volume, and computation of chaotic regions or magnetic islands takes considerably more effort~\citep{hudson2012}.

Accuracy of numerical \gls{PDE} solutions is inherently connected to the representation which defines gradients, and commonly used ideal \gls{MHD} equilibrium solvers with nested magnetic field topology can be differentiated accordingly:
A widely used finite-difference solver employed in the design of currently operating stellarator devices is \texttt{VMEC}~\citep{Hirshman}, another pseudo spectral solver is \texttt{DESC}~\citep{Dudt2020} and a third example is \texttt{GVEC}~\citep{Hindenlang2025}, that abstracts the notion of basis functions, which enabled computation of plasmas with figure-8 shape~\citep{Plunk2025}.\newline
Active control of stellarator plasmas is much less required than active control of tokamaks which are prone to disruptive events that can damage the machine because confinement in tokamaks is dependent on large toroidal plasma currents~\citep{Schissel2025}.
Modern control policies enabled accurate tracking of location, current and shape of axisymmetric plasmas realizable within the Tokamak à Configuration Variable device~\citep{Degrave2022}.
This shows that digital twins and real-time control can also be helpful tools in future stellarator devices, especially regarding control of transport and turbulence and possibly accessing novel plasma states by careful search in a devices' configuration space.
Many transport and turbulence codes, and accordingly their surrogate models, rely on either a coordinate system in which the magnetic field lines are straight~\citep{Mandell2024} or the equilibrium magnetic field~\citep{Landreman2014}.
Computation of straight field line coordinate systems requires the equilibrium magnetic field and models with very rapid inference of the equilibrium field will be helpful in sophisticated stellarator control strategies.
Furthermore, real-time interpretation of diagnostic data is facilitated if inference time of magnetic equilibria is reduced as much as possible~\citep{Merlo2023a}.

\Glspl{NN} enable quick inference by transferring the bulk of computation to training the \gls{NN}, which is then composed of simple non-linearities and parallelizable matrix multiplications.

We introduce simple \gls{NN}-based models with low residuals over parametrised spaces of equilibria within fixed-boundaries and with fixed rotational transform.
These models are parametrised by a unit interval scalar multiplier of the pressure coefficients and achieve volume-average force residuals very close to \descs force residual over the whole interval, from near-vacuum conditions to full pressure.

\subsection{Motivation}\label{sec:motivation}

This work takes the next step on the path to precise operator models of a subset of fusion relevant ideal \gls{MHD} equilibria by integrating \glspl{NN} into \desc.
Previous work presented advantages of small \glspl{MLP} which output Fourier decomposed equilibrium magnetic fields~\citep{Thun2026} and we test the same approach in \descs Fourier Zernike basis.

\desc can solve current and iota prescribed equilibria, includes many features such as omnigeneous field optimisation~\citep{Dudt2024}, mercier stability~\citep{Panici2023} and has implemented interfaces to gyrokinetic turbulence codes~\citep{Kim2024} - all this is immediately available to evaluate operator models parametrised by \glspl{NN} in future work.
The implementation of \desc allows for easy integration of \glspl{NN} and \descs optimisation subspace, in which linear constraints are satisfied by construction, reduces the dimensionality of the minimisation problem.
We train narrow operator models in \descs optimisation subspace ($\yprojected$ in equation~\eqref{eqn:desc_projection}) using only the force residual evaluated on typical concentric grids at discrete multipliers of the pressure coefficients.

Operator models that parametrise equilibria with low normalized force error are the scaffolding for digital twins, real-time control algorithms and rapid interpretation of diagnostic data.
Furthermore, precise equilibrium operator models of the configuration space of a machine are necessary to create sophisticated real-time capable simulation frameworks, for example, including transport and turbulence operator models which use deviations from the equilibrium magnetic field~\citep{Schissel2025}.

Another application for the presented operator models is in optimisation: Parametrised operator models ensure low sensitivity of stellarator optimisation targets towards uncertainty in the prescribed pressure profile.
The presented models are a first step towards parametrisation of \textit{flexible configurations} that preserve low optimisation metrics throughout a device's operational limits and map out the landscape of said metrics.

In terms of flight simulators or digital twins, control is likely to benefit from such models: Once trained, they can better inform control algorithms by rapidly propagating aleatoric uncertainties through the magnetic field topology to the control algorithm.
Models with rapid inference of plasma evolution are expected to play an important role in sophisticated control strategies of advanced fusion experiments~\citep{Schissel2025}.

Prior research determined that the minimum size, or complexity, of simple \glspl{MLP} which parametrise a single ideal \gls{MHD} equilibrium in Fourier space necessitate two hidden layers, each with a nonlinear activation function~\citep{Thun2026}.
\added{In the following, we will use this result to answer whether similar two-layer \glspl{MLP} are also capable of reproducing narrow regions of the ideal \gls{MHD} \gls{PDE} operator defined by a scaling factor of the pressure, while keeping the plasma boundary and the rotational transform profile constant.
Because we only vary the pressure in this work, we will call these models \emph{narrow} operator models in the sense that they parametrise a narrow subspace of the ideal \gls{MHD} \gls{PDE}.}

\section{3D ideal Magnetohydrostatic problem}\label{subsec:3d_mhd}
Stationary points of the ideal \gls{MHD} \gls{PDE} with isotropic pressure $p$ describe plasma as fluids with one species only in the limit of long-wavelengths, low-frequencies and no electric resistivity~\citep{Freidberg2014}
\begin{align}
    \mathbf{J} \times \mathbf{B} &= \pmb{\nabla} p \label{eqn:momentum}\\
    \mu_\mathrm{0} \mathbf{J} &= \pmb{\nabla} \times \mathbf{B} \label{eqn:ampere}\\
    \pmb{\nabla} \pmb{\cdot} \mathbf{B} &= 0  \label{eqn:gauss}
\end{align}
with the vacuum permeability $\mu_\mathrm{0}$.
Inserting Amp\`ere's law~\eqref{eqn:ampere} into the momentum equation~\eqref{eqn:momentum} removes currents $\mathbf{J}$ from this system of equations, yielding the residual force $\mathbf{F}$
\begin{align}
(\pmb{\nabla} \times \mathbf{B}) \times \mathbf{B} &= \mu_\mathrm{0} \, \added{\pmb{\nabla}} p \nonumber\\
\Leftrightarrow  \mathbf{F} &= (\pmb{\nabla} \times \mathbf{B}) \times \mathbf{B} - \mu_\mathrm{0} \, \added{\pmb{\nabla}} p\nonumber\\
\Leftrightarrow  \mathbf{F} &= F_\rho \pmb{\nabla}\rho + F_\theta \pmb{\nabla} \theta + F_\theta \pmb{\nabla} \zeta. \label{eqn:mhd_force}
\end{align}
Equilibrium states are defined by the \replaced{$\mathbf{B}$-field topology}{topology of the magnetic $\mathbf{B}$-field}, which has toroidal, or ring-shaped, form for magnetically confined plasmas in tokamaks and stellarators.

Under the assumption of nested, or integrable, structure of this magnetic field, the component in radial direction $\rho$ of the magnetic field $B^\rho=\mathbf{B}\pmb{\cdot} \pmb{\nabla} \rho$ is $0$, and we can write the magnetic field as
\begin{align}
    \label{eqn:b_f}
    \mathbf{B} &= \pmb{\nabla} \zeta \times \pmb{\nabla}  \chi + \pmb{\nabla} \psi \times \pmb{\nabla} \theta^{\star}\\
    &= B^\theta \mathbf{e}_\theta + B^\zeta \mathbf{e}_{\zeta} \nonumber
\end{align}
with toroidal magnetic flux $2\pi \psi$ and poloidal magnetic flux $2\pi \chi$.
The radial magnetic coordinate in this work is the same as \descs $\rho=\sqrt{s}=\sqrt {\psi/\psi_\mathrm{b}} \in \mathbb{R}\cap [0,1)$, $\theta^\star$ is a poloidal angle which straightens magnetic field lines and the magnetic toroidal angle $\zeta$ is equal to the cylindrical toroidal angle~\citep{Helander2014}.

Nestedness of the magnetic topology implies constant toroidal and poloidal magnetic flux on isobaric flux surfaces.
Assuming nested flux surfaces, the ideal \gls{MHD} equilibrium equations can be solved in an inverse manner, i.e.\ they are fully defined by the map from independent to cylindrical lab coordinates $[\rho, \theta, \zeta]^\mathsf{T}  \rightarrow [R, \lambda, Z]^\mathsf{T}$ and three invariants~\citep{Hirshman}.\newline
Under Gauss's law for magnetism, the contravariant $\mathbf{B}$-field reduces to
\begin{align}
    \mathbf{B}=\, \frac{\partial_\rho \psi}{\sqrt{g}} \left((\iota(\rho) \, - \, \partial_\zeta \lambda) \mathbf{e}_{\theta} +  (1+\partial_\theta \lambda) \mathbf{e}_{\zeta}\right).
    \label{eqn:b_contra}
\end{align}
Because the poloidal angle is arbitrary, as long as it is periodic and the Jacobian of the inverse map stays finite and does not switch sign, $\lambda$ is introduced as a renormalization function which straightens magnetic field lines: $\theta^{\star} = \theta + \lambda(\rho,\theta,\zeta)$.\newline
\citet{Hirshman} defined the covariant basis vectors of the inverse map as
$$
\mathbf{e}_\rho=\left[\begin{array}{c}
                       \partial_\rho R \\
                       0            \\
                       \partial_\rho Z
\end{array}\right] \quad \mathbf{e}_{\theta}=\left[\begin{array}{c}
                                                       \partial_{\theta} R \\
                                                       0                   \\
                                                       \partial_{\theta} Z
\end{array}\right] \quad \mathbf{e}_{\zeta}=\left[\begin{array}{c}
                                                      \partial_{\zeta} R \\
                                                      R                  \\
                                                      \partial_{\zeta} Z
\end{array}\right]
\label{eqn:basis_vectors}
$$
in conjunction with the Jacobian
\begin{equation}
    \sqrt{g}=\mathbf{e}_s \pmb{\cdot} \mathbf{e}_\theta \times \mathbf{e}_\zeta=(\mathbf{e}^s \pmb{\cdot} \mathbf{e}^{\theta} \times \mathbf{e}^\zeta)^{-1}.\label{eqn:jacobian}
\end{equation}
And the contravariant basis vectors are $\mathbf{e}^i = \pmb{\nabla} i = \frac{\mathbf{e}_j \times \mathbf{e}_k}{\sqrt{g}}$ with $(i, j, k)$ a cyclic permutation in $\{\rho, \theta, \zeta\}$.

The last components required for solving equations~\eqref{eqn:momentum} to~\eqref{eqn:gauss} are three invariants:
Any equilibrium needs some prescribed \added{(isotropic)} pressure profile $p(\rho)$, a rotational transform profile $\iota(\rho)$ or some current profile $c(\rho)$~\citep{Hirshman1985}, and optionally, the plasma boundary can be enforced via $R(\rho=1)=R_\mathrm{b}$ and $Z(\rho=1)=Z_\mathrm{b}$, in which case the equilibrium is called \textit{fixed-boundary}~\citep{Kruskal1958}.
\added{Additionally, the total toroidal flux through the torus $\psi_\mathrm{b}$ can be either specified or set to $1 \, \si{Wb}$ with later rescaling of other values.}

Finally, the output is an equilibrium magnetic field $\mathbf{B}$, determined by a balance between plasma pressure gradient and Lorentz force under the assumptions of nested magnetic surfaces inside a fixed plasma boundary.
The relative strength of both forces is commonly described by a ratio, the \textit{plasma beta}
\begin{equation}
    \betavol = \frac{\langle p \rangle_\mathrm{vol} \, 2 \mu_\mathrm{0}}{\langle \mathbf{B}^2 \rangle_\mathrm{vol}}
\end{equation}
with brackets denoting the volume average of some quantity $(\pmb{\cdot})$
\begin{equation}
    \langle (\pmb{\cdot}) \rangle_\mathrm{vol} = \frac{1}{V} \int_\rho \int_\theta \int_\zeta (\pmb{\cdot}) \sqrt{g} \, d\rho \, d\theta \, d\zeta.
    \label{eqn:volume_average}
\end{equation}
The plasma volume is computed by integrating the Jacobian~\eqref{eqn:jacobian} over the triplet $(\rho,\theta,\zeta)$.

Minimisation of equation~\eqref{eqn:mhd_force} is simplified by inserting the contravariant $\mathbf{B}$-field~\eqref{eqn:b_contra}\deleted{ into it}, \replaced{revealing}{which reveals} two independent directions of the covariant force: \added{On the one hand $F_\rho=\sqrt{g}(J^\zeta B^\theta - J^\theta B^\zeta) + \partial_\rho p(\rho)$ in radial direction $\pmb{\nabla} \rho$ and, on the other hand, $F_{\beta}=\sqrt{g}J^\rho$ in helical direction $\pmb{\beta}_\mathrm{DESC}=B^\zeta \pmb{\nabla} \theta - B^\theta \pmb{\nabla} \zeta$ ~\citep{Panici2023}}
\begin{align}
    \mathbf{F} = \left(\sqrt{g}(J^\zeta B^\theta - J^\theta B^\zeta) + \partial_\rho p(\rho) \right) \pmb{\nabla} \rho &+ \sqrt{g}J^\rho (B^\zeta \pmb{\nabla} \theta - B^\theta \pmb{\nabla} \zeta).\label{eqn:F_minimiser}
\end{align}
The currents in \deleted{this expression} are given by $\mu_0 J^i = \pmb{\nabla}\pmb{\cdot}(\mathbf{B}\times \pmb{\nabla} i)$ with $i$ a cyclic permutation of $\{\rho,\theta,\zeta\}$.

Numerical solutions to equations~\eqref{eqn:momentum} to~\eqref{eqn:gauss} \replaced{with different characteristics}{of different solvers} can be compared using the normalized force
\begin{equation}
    \mathbf{F}_\mathrm{norm} = \frac{|(\pmb{\nabla} \times \mathbf{B}) \times \mathbf{B} - \mu_\mathrm{0} \, \added{\pmb{\nabla}} p(\rho)|}{\,\,\langle |\pmb{\nabla} |B|^{2}/(2\mu_\mathrm{0})| \rangle_\mathrm{vol}}.
    \label{eqn:f_abs_norm}
\end{equation}
The denominator in this equation is the volume average of the magnetic pressure gradient with $\pmb{\nabla} |B|^2 = 2  (|B|  \pmb{\nabla} |B|)$.
We will denote the scalar volume average of $\mathbf{F}_\mathrm{norm}$ as $\fvolnorm$ in the following (see e.g.\ figure~\ref{fig:fvol_all}).

Due to physics and engineering reasons, stellarators are commonly split into $\nfp$ self-similar parts, each occupying $2\pi / \nfp$ of the full toroidal angle $\zeta\in[0, 2\pi)$.
\subsection{\texttt{DESC} solver}\label{sec:desc}
\desc is a pseudo spectral code that not only efficiently solves equations~\eqref{eqn:momentum} to~\eqref{eqn:gauss}, but also includes other important stellarator minimisation problems.
It can solve free-boundary equilibria where, instead of fixing the plasmas boundary, a current is specified some distance from the plasma boundary which is then split into discrete coils~\citep{Conlin2024a}.
\desc implements stellarator optimisation targets, such as the direct minimisation of omnigeneous field errors, Mercier or infinite-$N$ ideal ballooning stability, and is coupled to other codes like the turbulence code \texttt{GX} or the gyrokinetic solver \texttt{GS2}~\citep{Gaur2024, Kim2024}.
The previously mentioned analytic near-axis expansion can be used as a starting point for optimisation in \desc, which then computes solutions valid throughout the volume.
A comparison between \desc, \texttt{VMEC} and a code which can resolve magnetic islands, \texttt{SPEC}, agreed well on the magnetic axis position of a Heliotron like equilibrium~\citep{Hudson2025}.

\desc minimises equation~\eqref{eqn:F_minimiser} by weighting $F_\rho$ and $F_\beta$ with the occupied volume of each collocation point:
\begin{align}
    f_{\rho} = F_{\rho} ||\pmb{\nabla} \rho||_2 \sqrt{g} \Delta \rho \Delta \theta \Delta \zeta \\
    f_{\beta} = F_{\beta} ||\pmb{\beta}_\mathrm{DESC}||_2 \sqrt{g} \Delta \rho \Delta \theta \Delta \zeta
\end{align}
A non-linear system of equations is then solved by least-squares optimisers
\begin{equation}
    \mathbf{f}(\mathbf{x}) = \left[\begin{array}{c}
    f_{\rho, j}(\mathbf{x}) \\
    f_{\beta, k}(\mathbf{x})
    \end{array}\right] = \mathbf{0}
\end{equation}
\added{for $\mathbf{x}=[\mathbf{R}_{lmn}(\rho, \theta, \zeta), \pmb{\lambda}_{lmn}(\rho, \theta, \zeta), \mathbf{Z}_{lmn}(\rho, \theta, \zeta)]^\mathsf{T}$} with $j=0, ..., J$ and $k= 0, ..., K$ indexing collocation points on possibly two different grids.
\replaced{$\mathbf{x}$ are the coefficients for each dependent coordinate decomposed into a Fourier-Zernike basis with Zernike polynomial of finite radial order $l=0, 1, ..., L$ and finite poloidal mode numbers $m=-M, -M+1, ..., 0, ..., M-1, M$, and a toroidal decomposition into Fourier modes with finite mode numbers $n=-N, -N+1, ..., 0, ..., N-1, N$. The radial function of the Zernike polynomials is a shifted Jacobi polynomial, which fulfills the mathematical condition for physical scalars on the unit disc~\citep{Lewis1990}. Due to brevity, we omit the full basis description here and refer interested readers to~\citet{Dudt2020} and~\citet{Panici2023}.}{\desc expands the independent coordinates $(\rho,\theta,\zeta)$ in Zernike polynomials for $(\rho, \theta)$, which fulfill the mathematical condition for physical scalars on the unit disc Lewis1990, and Fourier transformation in $\zeta$. Dudt2020 provides a more precise mathematical description of the expansions in DESC.}

Each equilibrium solved by \desc is defined by the result of the \added{non-linear} least squares minimisation
\begin{align}
    \text{minimize} \hspace{1cm}    &||(\pmb{\nabla} \times \mathbf{B}) \times \mathbf{B} - \mu_\mathrm{0} \, \pmb{\nabla} p(\rho)||_2^2 \label{eqn:desc_min_prob}  \\
    \text{subject to} \hspace{1cm}    &\mathbf{A\bar{x}=b}\nonumber
\end{align}
where $\mathbf{b}$ is the target for the constraints in $\mathbf{\bar{x}}=[\mathbf{x}, \mathbf{c}]$.
In this work, the constraint vector includes the ideal \gls{MHD} invariants for fixed-boundary equilibria, namely $\mathbf{c}=[\mathbf{R}_{b,mn}, \mathbf{Z}_{b,mn}, \mathbf{p}_l, \pmb{\iota}_l, \psi_\mathrm{b}]^\mathsf{T}$ ~\citep{Conlin2022}. $\pmb{\iota}_l$ are the coefficients of the prescribed rotational transform profile and $\mathbf{p}_l$ are the coefficients of some isotropic pressure function (see table~\ref{tab:hyperparams}).

The constrained minimisation problem defined by~\eqref{eqn:desc_min_prob} is then transformed into an unconstrained problem by splitting $\mathbf{\bar{x}}$ into a particular solution $\mathbf{x}_\mathrm{p}$ and a vector $\mathbf{y}$ on a hyperplane defined by the constraint manifold $\mathbf{A} \mathbf{x} = \mathbf{b}$.
\begin{align}
    &\mathbf{A}\mathbf{Z} = 0 \nonumber \\
    &\mathbf{A} \mathbf{\bar{x}} = \mathbf{A}(\mathbf{x}_\mathrm{p} + \mathbf{Z}\mathbf{y}) = \mathbf{b} \nonumber \\
    \iff &\mathbf{A} (\mathbf{x}_\mathrm{p} + \mathbf{Z} \mathbf{y}) = \mathbf{b}.\label{eqn:desc_projection}
\end{align}
The nullspace $\mathbf{Z}$ then only needs to be computed once before the start of optimisation via singular value decomposition, yielding an unconstrained optimisation problem over the projected parameter vector $\mathbf{y}=\mathbf{Z}^\mathsf{T}(\mathbf{\bar{x}}-\mathbf{x}_\mathrm{p})$.

\desc commonly uses an incremental minimisation, or \emph{automatic continuation}, parametrised by two multipliers: $\eta_\mathrm{b}\in[0, 1]$ and $\etap \in [0, 1]$.
Starting with a circular Tokamak, each incremental step either changes $\eta_\mathrm{b}$, which deforms the initial circular plasma boundary into the desired stellarator shape, or $\etap$, which ramps up the pressure.

The \gls{NN} based parametrisation of this work sets $\mathbf{y}$ as the output layer of simple \glspl{MLP} and then minimises the sum of force residuals of equilibria defined by a set $\etaptrain$.

\section{Physics informed Neural Networks in \desc}\label{subsec:pinns}

Instead of directly optimising over $\yprojected$ \added{(equation~\eqref{eqn:desc_projection})} as done in \desc, we will set $\yprojected$ as the output of two-layer \glspl{MLP} and optimise over the \gls{MLP}'s parameters $\bm{\nu}$ in a \gls{PINN} approach~\citep{Raissi2019}.
To this end, we reformulate the optimisation problem as
\begin{align}
\bm{\nu}^{\star} = \argmin_{\bm{\nu}} \quad \mathcal{L}_{\mathrm{op}}
    \label{eq:pinn_general}
\end{align}
and solve it with \texttt{L-BFGS} optimiser for some loss function $\mathcal{L}_{\mathrm{op}}$.
\citet{Paluzo-Hidalgo2020} showed that \glspl{MLP} with two hidden layers and non-linearities consisting of rectified linear units can approximate arbitrary functions, and in Fourier space, optimising over parameters of \glspl{MLP} with two hidden layers is sufficient to solve single, fixed-boundary and finite-$\betavol$ \gls{MHD} equilibria with lowest $\fvolnorm$ \added{over the tested spectral resolutions and \glspl{NN}}~\citep{Thun2026}.
Setting the number of hidden layers or their node numbers too large in the presented approach yields unusable results because minimisation stagnates in local optima.

The plasma boundary is fixed via projection into $\yprojected$ (see equation~\eqref{eqn:desc_projection}), which reduces the number of parameters and, in our tests, optimising \gls{NN} with the Fourier Zernike modes in the output layer did not work.
As initial guess for all presented narrow operator models we use the default \texttt{DESC} initial guess.
If it fails to produce nested flux surfaces, for example for the quasi-helical equilibrium (figure~\ref{fig:qs_qh}), an invertible mapping to boundary conforming coordinates introduced by~\citet{Babin2025} calculates the $N=2$ axis initial guess.
This axis guess is then interpolated towards the boundary, ensuring a well posed initial guess throughout the volume.
The initial guess in Fourier Zernike space is projected into the tangent space as $\mathbf{y}_\mathrm{init}$, \added{using the same $\mathbf{A}$ and $\mathbf{Z}$ matrices}, and then added to the \gls{MLP} prediction (see equation~\eqref{eqn:full_prediction}).

We show that it is possible to minimise the sum of ideal \gls{MHD} force residuals defined by equilibria evenly distributed in $\eta_\mathrm{p, train}$ over the parameters of \glspl{MLP} with two hidden layers.
Each operator \gls{MLP} parametrises the function $\text{MLP}: \eta_{\mathrm{p}, i} \rightarrow \yprojected_i$ for $\etaptrain=\{\eta_{\mathrm{p},i=0},...,\eta_{\mathrm{p},i=I-1}\}$.
The input can be easily modified to include, for example, the boundary Fourier modes or rotational transform coefficients, but results in this work only use the scalars $\eta_{\mathrm{p}, i}$ as input.

Each full training step consists of the model predicting $\mathbf{y}_\mathrm{train}$ for all $\eta_{\mathrm{p}, i}$ and the sum of all residuals for all $i$ as target function, scaled by $\alpha_\mathrm{MHD}=10^7$ to avoid optimisation problems caused by the residual approaching machine precision
\begin{equation}
    \mathcal{L}_{\mathrm{op}} = \alpha_\mathrm{MHD} \sum_{i=0}^{I-1} |\mathbf{f}(\mathbf{x})|^2_i =  \alpha_\mathrm{MHD} \sum_{i=0}^{I-1} |\hat{\mathbf{f}}(\mathbf{y})|^2_i \label{eqn:loss}
\end{equation}
where $\hat{\mathbf{f}}$ is the composition of $\mathbf{f}$ and the inverse of the linear projection \added{$\mathbf{\bar{x}} = \mathbf{x}_\mathrm{p} + \mathbf{Z} \mathbf{y}$}.

We use $I=10$ equispaced $\eta_{\mathrm{p, train}, i}$ points to train all presented narrow operator models.
The \glspl{MLP} use the same activation function as self-normalizing \glspl{NN}~\citep{Klambauer2017}, which~\citet{Merlo2021} also deemed optimal through hyperparameter search
\begin{equation}
     \begin{split}\sigma(x) =\mathrm{selu}(x) = \lambda_s \begin{cases}
  x, & x > 0\\
  \alpha_s e^x - \alpha_s, & x \le 0
\end{cases}\end{split}
    \label{eqn:selu}
\end{equation}
with $\lambda_s=1.0507...$ and $\alpha_s=1.6732...$\,.\newline
Each \gls{MLP} has the following functional form
\begin{align}
\hat{\mathbf{y}}_\mathrm{mlp}(\etaptrain) &= \mathbf{W}_\mathrm{2}(\sigma \mathbf{z}_1(\etaptrain)) + \mathbf{b}_\mathrm{2}\\
\mathbf{z}_\mathrm{1}(\etaptrain) &= \mathbf{W}_\mathrm{1}(\sigma \mathbf{z}_0(\etaptrain)) + \mathbf{b}_\mathrm{1}\\
\mathbf{z}_\mathrm{0}(\etaptrain) &= \mathbf{W}_\mathrm{0}(\etaptrain) + \mathbf{b}_\mathrm{0}.
\end{align}
Weights of the \glspl{MLP} $\mathbf{W}_l$ for $l=0, 1, 2$ are initialised with a normal distribution $\pmb{\mathcal{N}}(0, 0.01^2)$, while the bias vectors $\mathbf{b}_l$ for $l=0, 1, 2$ are initialised with $\mathbf{0}$.

The \gls{MLP} output is scaled and added to the linear projection of the initial guess, which is necessary for convergence for all non-axisymmetric equilibria we tested:
\begin{equation}
    \yprojected = \mathbf{y}_\mathrm{init} + \mathbf{y}_\mathrm{scale} \hat{\mathbf{y}}_\mathrm{mlp}.\label{eqn:full_prediction}
\end{equation}
$\mathbf{y}_\mathrm{init}$ is the projected initial guess, and the scaling vector $\mathbf{y}_\mathrm{scale}$ is the projection of the inverse of the sum of absolute mode numbers $l, m$ and $n$ (see table~\ref{tab:hyperparams} for the non-projected scales).
All $\mathbf{y}$ in~\eqref{eqn:full_prediction} are projected with the same $\mathbf{A}$ and $\mathbf{Z}$ operators \added{(see equation~\eqref{eqn:desc_projection})}.

Optimisation of each \gls{MLP} is split into two stages: First, the loss function is modified to only include the outliers,~i.e. $i=0$ and $i=I-1$, and in a second minimisation all $I$ equilibria are included.
The trained models are then tested on $\etaptest$ which oversamples $I$ by a factor of $10$, staying within the interval $[\eta_{\mathrm{p},0}, \eta_{\mathrm{p},I-1}]$ (see figure~\ref{fig:fvol_all}) and an extrapolation of each model is plotted in figure~\ref{fig:extrap}.
We provide detailed hyperparameters for models and optimisation in table~\ref{tab:hyperparams} and code which reproduces the plots in the supplementary data (see Appendix~\ref{asec:data}).

\section{Results}
\label{sec:results}
This section compares \descs \texttt{lsq-exact} optimiser with an \texttt{L-BFGS} optimiser applied to the free parameters $\pmb{\nu}$ of \glspl{MLP} that parametrise the linear projection of the Fourier Zernike basis over $\etaptrain$.
All equilibria we show are fixed-boundary equilibria with $\betavol>0$ and constant rotational transform or current profile.
Furthermore, all results presented in this section do not use continuation methods or iterative refinement of the grid and compute $\mathcal{L}_\mathrm{op}$ (equation~\ref{eqn:loss}) on concentric grids commonly used by \desc~\citep{Conlin2022}.
Volume- or surface-averaged quantities are calculated on quadrature grids.

Each \desc equilibrium in this comparison is solved in two stages: First, the equilibrium is optimised with automatic continuation and moderate tolerances, and in a second optimisation the tolerances of the resulting equilibrium are reduced to zero with a maximum of $100$ iterations.
If automatic continuation yields intermediate equilibria that \desc cannot solve, we instead solve the equilibrium without automatic continuation.
This is only the case for some $\eta_{\mathrm{p},i}$ in the quasi-helical configuration.
Automatic continuation where $\etap$ is increased first can fail due to intermediate equilibria having unrealistic pressure, and this will be remedied in a future version of \desc by performing continuation in $\eta_\mathrm{b}$ first and then $\etap$.
Lastly, we run \desc with the same spectral resolutions $M$ and $N$ as prescribed in the input files which are included in the supplementary material, and the Zernike polynomials are of order $L=M$.
Except for the \gls{W7X} equilibrium where we use $L=M+1=7$ as $L=6$ did not resolve the \replaced{S}{s}hafranov shift properly.

Comparing the \gls{MLP} operator model to \desc solutions is only possible at discrete points \added{$\etaptrain$} due to \desc solving single instead of spaces of equilibria.
\desc solutions in figure~\ref{fig:fvol_all} are marked with a plus sign while the training points of the operator models are marked by a cross.

We present operator models for a $\nfp=5$ \gls{W7X}-like equilibrium in standard configuration, a $\nfp=19$ Heliotron-like equilibrium, a $\nfp=4$ quasi-helical equilibrium and an axisymmetric, but not stellarator symmetric, equilibrium akin to the experimental device DIII-D.
The \gls{W7X} equilibrium is a good example for three-dimensional plasma in an experimental device, and the quasi-helical equilibrium is representative for optimised quasisymmetric stellarators with self-consistent bootstrap current~\citep{Landreman2022a}.

Out of all three stellarator equilibria, the Heliotron-like equilibrium has the highest sensitivity of its axis position with respect to the plasma $\betavol$, moving its axis by $25.7\si{cm}$ between $\betavol(\eta_{\mathrm{p},0})=1.018\%$ and $\betavol(\eta_{\mathrm{p},9})=10.18\%$.
The operator model is able to resolve this change along $\eta_{\mathrm{p},i}$ as seen in figure~\ref{fig:heliotron_op}.
Additionally, we plot the \replaced{flux surfaces}{solution} of operator \replaced{\gls{NN}}{models} similar to DIII-D and \gls{W7X} in standard configuration in Appendix~\ref{asec:w7x_d3d}.

Table~\ref{tab:hyperparams} provides the optimisation parameters of each narrow operator model.
The last line of table~\ref{tab:hyperparams} shows the maximum discrepancy for $\etaptrain$ between the \desc solution and the model solution
\begin{equation}
    \epsilon_\mathrm{X} = \frac{1}{N} \sum_i \sqrt{(R_{NN,i} - R_{\desc,i})^2 + (Z_{NN,i} - Z_{\desc,i})^2}\label{eqn:epsx}
\end{equation}
evaluated on equal grid with collocation points indexed by $i=0, 1, ..., N$.

\begin{table*}
    \begin{center}
        \captionsetup{width=0.8\textwidth}
        \def~{\hphantom{0}}
        \begin{tabular}{lcccc}
        \toprule
        \textbf{Parameter} & \textbf{DIII-D} & \textbf{Heliotron} & \textbf{W7-X} & \textbf{Quasi-helical} \\
        \midrule
        \gls{NN} nodes per layer          & $1,8,16,182$    & $1,64,256,351$ & $1, 32, 64, 265$ & $1, 32, 128, 661$   \\
        $M$                         & $12$ & $9$ & $6$ & $8$ \\
        $N$                         & $0$ & $3$ & $6$ & $8$ \\
        $x_{\mathrm{scale},j}$  & $1$ & $\frac{1}{l_j + |m_j| + |n_j| + 1}$   & $\frac{1}{l_j + |m_j| + |n_j| + 1}$ & $\frac{1}{l_j + |m_j| + |n_j| + 1}$   \\
        current prescribed      & yes   & no    & no   & yes    \\
        $\betavol$ at $\eta_\mathrm{p} = 1$ & $0.645\%$  & $10.18\%$  & $4.43\%$  & $5.08\%$   \\
        pressure function type & $\displaystyle \sum_{m} \rho^{2m} a_m$ & $\displaystyle \sum_m \rho^{2m} a_m$ & $\displaystyle \sum_m \rho^{2m} a_m$ & $\displaystyle \sum_m a_m B_m^3(\rho^2) $ \\
        $\displaystyle \max_{\etaptrain}(\epsilon_\mathrm{X}(\etap)) \, [\si{m}]$  & $\epsilon_\mathrm{X}(1.0)=0.0062$ & $\epsilon_\mathrm{X}(1.0)=0.0079$ & $\epsilon_\mathrm{X}(0.1)=0.0134$ & $\epsilon_\mathrm{X}(0.4)=0.06$ \\
        \bottomrule
        \end{tabular}
        \caption{Summary of model and equilibrium parameters. The last number of nodes per layer is the size of $\yprojected$ and $B_m^3$ is a cubic B-spline basis.}
        \label{tab:hyperparams}
        \end{center}
\end{table*}
Figure~\ref{fig:heliotron_op} illustrates the $\mathbf{B}$-field topology at $\zeta=0$ of the Heliotron equilibrium for $\eta_\mathrm{p}=\{0.21, 0.55, 0.89\}$, which are all points on which the model was not trained, but which lie within the training \added{interval $\etaptrain$}.
The \gls{MLP} parametrised solution is plotted in red while the solution of \texttt{DESC} is plotted in green and both agree well for all $\etap \in [0.1, 1]$.

Figure~\ref{fig:fvol_all} plots the scalar quantity $\fvolnorm$ over $\etap \in [0.1, 1]$ for all tested equilibria.
The models and \texttt{DESC}'s force error of the Heliotron equilibrium match well for all $\eta_\mathrm{p} \in [0.1, 1]$ with the largest discrepancy around $\eta_p=0.11$, where the operator model shows a small spike in force error.
Removing this spike requires increasing the number of training points $I$.
All \gls{NN} models show good agreement with \texttt{DESC} and stay below $\fvolnorm<1\%$.
In contrast, \texttt{DESC} achieves comparable $\fvolnorm$ in the quasi-helical and lower $\fvolnorm$ for the DIII-D and \gls{W7X} test cases for all $\etaptrain$.

To illustrate quantities of interest that depend on higher order derivatives, we showcase quasi-symmetry in Boozer coordinates for the quasi-helical equilibrium with self-consistent bootstrap current computed at $\etaptrain=1$.
Throughout $\etaptest$, the quasi-helical operator \gls{MLP} shows good quasi-symmetry at radial position $s=\rho^2=0.75$ plotted for $\etap=\{0.21, 0.55, 0.89\}$ in figure~\ref{fig:qs_qh}.
Only the maxima of $|\mathbf{B}|$ change slightly with decreasing $\etap$ at $\theta_\mathrm{Boozer}$ close to $0$.
The topology of the magnetic well is qualitatively preserved.
\begin{figure*}
    \captionsetup{width=0.9\textwidth}
    \centering
    \resizebox{\textwidth}{!}{\input{fig1.pgf}}
    \caption{Operator \gls{MLP} solution of Heliotron-like equilibria at $\zeta=0$ with elliptical boundary for different pressure scaling factors $\eta_\mathrm{p}$. The \texttt{DESC} solution in green and \gls{MLP} solution in red match qualitatively for the plotted flux surfaces.}\label{fig:heliotron_op}
\end{figure*}
\begin{figure}
    \centering
    \resizebox{0.8\textwidth}{!}{\input{fig2.pgf}}
    \vspace{2em}
    \caption{Operator \gls{MLP} solutions for equilibria presented in this work and trained on $I=10$ equispaced $\etaptrain$ (cross signs) compared to their \texttt{DESC} solution (plus signs) at the training points in terms of $\fvolnorm$.}\label{fig:fvol_all}
\end{figure}
\begin{figure*}
    \centering
    \captionsetup{width=0.85\textwidth}
    \resizebox{\textwidth}{!}{\input{fig3.pgf}}
    \caption{Good quasisymmetry for the $\nfp=4$ quasi-helical equilibrium at $s=\rho^2=0.75$ for $\etap=\{0.21, 0.55, 0.89\}$ with a constant current that was optimised for $\etap=1$. $\theta_\mathrm{Boozer}$ and $\zeta_\mathrm{Boozer}$ are straight-fieldline coordinates in which transport equations are nearly isomorphic to axisymmetric equilibria~\citep{Pytte1981}.}\label{fig:qs_qh}
\end{figure*}

\subsection{Discussion}\label{sec:discussion}

Optimisation of the presented narrow operator learning models was stopped at an arbitrary number of iterations, and further optimisation could yield models that close existing gaps between \texttt{DESC}'s and the model's $\langle \mathbf{F} \rangle_\mathrm{vol, norm}$ (see figure~\ref{fig:fvol_all}).

\added{Including $\etap<0.1$ in the training set resulted in slight differences of the $\mathbf{B}$-field topology between the model and \texttt{DESC} solutions at low $\etap$.
This could be caused by large scale differences in $\yprojected$ and whether such low beta regions are relevant for flight simulators remains an open question.
Also, rigid start-up sequences of experimental plasmas~\citep{Grulke2024} and control at high densities increase the importance of operator models closer to $\etap=1$.}
Evaluating such models at $\eta_\mathrm{p, test}>1$,~i.e. outside $\eta_\mathrm{p, train}$, shows a monotonic increase in $\langle \mathbf{F} \rangle_\mathrm{vol, norm}$ (see Appendix~\ref{asec:extrap}).
Extrapolation with this approach to unseen equilibria seems unlikely, but extending $\etaptrain$ to relevant regions is straightforward.

Increasing the \gls{MLP} layer-size or -depth, or increasing the spectral resolution $L$, $M$ or $N$ too much, forces the minimisation to settle in local minima, far away from \texttt{DESC}'s optima.
Automatic continuation methods similar to \texttt{DESC} could avoid these local minima.

Training of simple two-layer \glspl{MLP} on single equilibria at $\etap=1$ without any continuation methods reduced $\fvolnorm$ close to \descs solutions, but not as low as~\citet{Thun2026} for \gls{W7X} and the Dshape Tokamak.
We do not include these results in this work due to brevity.

Over all, the presented optimisation of narrow operator models yields models that capture the equilibria in $\etaptest$ as good as \texttt{DESC} and can even achieve lower $\fvolnorm$ in some regions.
Here, at least in the tested cases, we see that optimising over an ensemble of equation~\eqref{eqn:desc_min_prob}, parametrised by the set $\eta_{\mathrm{p, train},i}$, can yield a continuous and  precise model of the narrow \gls{PDE} operator.
To verify this, we optimised $\yprojected$ with \desc again until the change in parameters was below machine precision (i.e.\ with a ceiling of $10^6$ \texttt{lsq-exact} iterations) and arrived at the same conclusion and qualitatively equal results as plotted in figure~\ref{fig:fvol_all}.

Quantities that depend on higher order derivatives of the equilibrium magnetic field such as the quasi-symmetry evaluated in Boozer coordinates and the magnetic well are preserved for the quasi-helical test case (figure~\ref{fig:qs_qh}).

Training these narrow operator models incurs higher computational costs compared to verifying the model with $I$ \desc solutions (see Appendix~\ref{asec:runtime}), however, the increased cost must be weighted against the advantages of continuously parametrised models over $\etaptest$.

One common approach to training ideal \gls{MHD} operator models is to construct a dataset of equilibrium magnetic fields with a conventional solver and then training a model on this dataset, and possibly an additional physics-based addition to the loss function.
For the quasi-helical equilibrium in figure~\ref{fig:fvol_all}, this training scheme would not improve upon \descs force error for $\etap<0.4$, hinting that additional training of operator models directly on the physics yields more precise models, \added{or optimisation potential in \descs minimisation for this specific equilibrium at $\etap<0.4$}.

\citet{Merlo2023} also presents improvements in optimisation with operator models trained on a surrogate for the force residual (equation~\eqref{eqn:desc_min_prob}) that assumes the helical force $F_\beta$ to be zero.

\subsection{Outlook}\label{sec:outlook}

To improve the applicability and training efficiency of the presented models, future work should explore the sampling granularity in $\etaptrain$ required to achieve good force error over the parameter range, increase $\pmb{\nu}$ and the \gls{NN} complexity while introducing more parameters like $\iota(\rho)$ coefficients.
A solution for the optimisation stagnating in local minima must be found when increasing the number of parameters.

In the Heliotron case, the model shows a spike between training points at $\etap \in [0.1, 0.2]$, whereas the other operator models follow a continuous trend, raising the question of how many training points $I$ are required for the latter without degradation of $\fvolnorm$ in $\etaptest$.
Investigating the optimal ratio of data from a solver and direct force residual in training sets of operator models could further reduce computational cost and help avoid local minima.

Finding commonalities in the parameters of the presented narrow operator models in terms of $\mathbf{x}$ instead of $\yprojected$ could yield more efficient optimisation, also named \textit{transfer learning} in \gls{PINN} research~\citep{Goswami2022}.
Care must be taken with regard to differently shaped $\yprojected$, but modification of the linear operators $\mathbf{A}$ and $\mathbf{Z}$ to include all equilibria under consideration can alleviate those issues.

Modification of the presented approach to include different inputs, for example the boundary coefficients, can improve sensitivity analysis because first order gradients of the force residual to the input space are easily computed by automatic differentiation.
Using one of the presented narrow operator models delivers continuous gradients of dependent to independent variables over the space parametrised by $\etaptest$.

In quasi-isodynamic stellarator optimisation the pressure profiles are usually fixed a-priori~\citep{Gaur2024, Sanchez2023, Goodman2024} and more diverse profiles could yield lower multi-objective targets or more flexible configurations.
Optimisation for flexible configurations could also yield more robust optimised stellarators.

Extending the narrow operator models to free-boundary equilibria is not straightforward:
The \desc suite already includes numerical free-boundary computation, but in our preliminary research we found that continuation methods are indispensable to solve free-boundary problems in \desc and these incur a change in the shape and encoded information of $\yprojected$.
However, reevaluating free-boundary operator models in \desc is more promising with the mentioned improvements in this section, especially transfer learning in terms of $\mathbf{x}$ and customized linear matrices $\mathbf{A}$ and $\mathbf{Z}$.

\section{Conclusion}\label{sec:final}
We presented narrow operator models in the form of \glspl{MLP} with two hidden layers parametrised by a scaling factor of the pressure coefficients.
These \glspl{MLP} reduce $\fvolnorm$ of various equilibria types to comparable levels computed by the modern solver \desc.
All introduced models precisely interpolate through an oversampled test set $\etaptest$ with all points still inside the training set $\etaptrain$ for all tested equilibria (see figure~\ref{fig:fvol_all}).
Importantly, they are trained only on a physics-based residual (equation~\ref{eqn:desc_min_prob}), that is without using any equilibrium information pre-computed by existing solver like \desc.

For the DIII-D like and \gls{W7X} equilibria, \desc computes equilibria with marginally lower $\fvolnorm$ compared to the \glspl{MLP} parametrisation, while for the quasi-helical equilibrium the \gls{MLP} approaches slightly lower $\fvolnorm$ than \desc for $\etap \in [0.4, 0.9]$ (see figure~\ref{fig:fvol_all}).
This is interesting for future operator model optimisation because it hints at a benefit when training on the force residual:
If the training set only consisted of pre-computed \desc solutions, the operator model's $\fvolnorm$ would be bounded by \descs $\fvolnorm$.

The narrow operator model of the quasi-helical equilibrium with self-consistent bootstrap current preserves good higher order metrics such as quasi-symmetry and magnetic well throughout $\etaptest$ (see figure~\ref{fig:qs_qh}).

Extrapolation of the model to unseen $\etap>1$ incurs a monotonically increasing $\fvolnorm$ (see Appendix~\ref{asec:extrap}), but including $\etaptrain > 1$ in the training set is straightforward.

Our training scheme was purposedly kept minimal, in the sense that no advanced enhancements from the \gls{PINN} literature were included, to evaluate simple \glspl{MLP} trained solely on the force residual as a baseline for future work.
We expect significant improvements to the presented method if recent advances in \gls{PINN} research are included~\citep{Luo2025}.

\newpage
\paragraph{\textbf{Acknowledgement}}

The authors thank M.~Landreman for valueable feedback and provision of the quasi-helical equilibrium, and we thank J.~Geiger for providing the \gls{W7X} equilibrium.
Our gratitude extends to E.~Kolemen for guidance throughout this work and international visiting opportunities.
Furthermore, we appreciate R.~Gaur's feedback during the project.
Last but not least, we are indebted to the various maintainers of the open source codes used in this work: \allowbreak \texttt{jax}, \allowbreak \texttt{optax}, \allowbreak \texttt{jaxopt}, \allowbreak \texttt{flax}, \allowbreak \texttt{orbax}, \allowbreak \texttt{desc} and \texttt{matplotlib}.
T.T.~is supported by a grant from the Simons Foundation (Grant No. 601966).
D.P.~is funded through the SciDAC program by the US Department of Energy, Office of Fusion Energy Science and Office of Advanced Scientific Computing Research under contract number DE-AC02-09CH11466, DE-SC0022005, and by the Simons Foundation/SFARI (560651).
This work was supported by a DOE Distinguished Scientist Award via DOE Contract DE-AC02-09CH11466 at the Princeton Plasma Physics Laboratory.
This work has been carried out within the framework of the EUROfusion Consortium, funded by the European Union via the Euratom Research and Training Programme (Grant Agreement No 101052200 - EUROfusion).
Views and opinions expressed are, however, those of the author(s) only and do not necessarily reflect those of the European Union or the European Commission.
Neither the European Union nor the European Commission can be held responsible for them.
\appendix
\section{\gls{W7X} and DIII-D equilibria}
\label{asec:w7x_d3d}
Here, we also show Poincar\`e sections for the DIII-D like and \gls{W7X} equilibria in figure~\ref{fig:d3d_poinc} and ~\ref{fig:w7x_poinc}.
The slight gap in $\fvolnorm$ between \desc and the \gls{NN} for low $\etap$ leads to a visible shift in the axis position for \gls{W7X} in standard configuration (figure~\ref{fig:w7x_poinc}, $\etap=0.21$).
\begin{center}
\captionsetup{width=0.9\textwidth}
\centering
\resizebox{0.9\textwidth}{!}{\input{fig4.pgf}}
\captionof{figure}{Operator \gls{MLP} solution of an axisymmetric equilibria akin to DIII-D and parametrised by pressure scaling factor $\eta_\mathrm{p}$. The \texttt{DESC} solution in green and \gls{MLP} solution in red match qualitatively for the plotted flux surfaces.}\label{fig:d3d_poinc}
\end{center}
\begin{figure}
    \captionsetup{width=0.9\textwidth}
    \centering
    \resizebox{0.9\textwidth}{!}{\input{fig5.pgf}}
    \caption{Operator \gls{MLP} solution of \gls{W7X} equilibria in standard configuration parametrised by pressure scaling factor $\eta_\mathrm{p}$. The \texttt{DESC} solution in green and \gls{MLP} solution in red match qualitatively for the plotted flux surfaces, except at $\etap=0.21$.}\label{fig:w7x_poinc}
\end{figure}
\section{Data availability}
\label{asec:data}
\desc version 13.0 was used for the results in this work.
Data, which includes \gls{MLP} parameters $\pmb{\nu}$, is provided at https://doi.org/10.5281/zenodo.17360265, including scripts that use trained $\pmb{\nu}$ to regenerate the plots of this work.
\section{Extrapolations}
\label{asec:extrap}
Extrapolation of a model trained on $\etaptrain \in [0.1, 1]$ to $\etaptest>1$ does incur a monotonically increasing force error, as seen in figure~\ref{fig:extrap}.
\begin{center}
    \resizebox{0.8\textwidth}{!}{\input{fig6.pgf}}
    \captionof{figure}{Extrapolation of all presented narrow operator models to $\etaptest>1$, outside of $\etaptrain \in [0.1, 1]$, shows a monotonic increase in $\fvolnorm$ with increasing $\etaptest$.}\label{fig:extrap}
\end{center}
\section{Run time}\label{asec:runtime}
All \desc and \gls{MLP} results were computed on the same machine.
The operator models for non-axisymmetric equilibria take roughly one to two orders of magnitude more compute resources to train compared to the $10$ \desc solutions, while the axisymmetric DIII-D like case was comparable in terms of compute resources to the $10$ \desc solutions.
We expect improvements to the \gls{MLP} approach if enhancements from the \gls{PINN} literature or automatic continuation are added to the current \gls{MLP} minimisation.
\section*{Declaration of Interest}
The authors report no conflict of interest.
\bibliographystyle{jpp}
\bibliography{jpp-instructions}
\end{document}